\documentclass[reprint,superscriptaddress,amsmath,amssymb,longbibliography,floatfix]{revtex4-2}

\usepackage{graphicx}
\usepackage{dcolumn}
\usepackage{bm}
\usepackage{hyperref}

\usepackage{booktabs}
\usepackage{multirow}
\usepackage{physics}
\usepackage{subcaption}
\usepackage{xcolor}

\newcommand{\etaGK}{\eta_\mathrm{GK}}
\newcommand{\etarh}{\eta_{\Rh}}
\newcommand{\Dgk}{D_\mathrm{GK}}
\newcommand{\Drh}{D_{\Rh}}
\newcommand{\Dpbc}{D_\mathrm{PBC}}

\newcommand{\scan}{SCAN}
\newcommand{\optb}{optB88-vdW}
\newcommand{\pbed}{PBE-D3}
\newcommand{\ff}{FF}            

\newcommand{\Lbox}{L_\mathrm{box}}      

\newcommand{\kB}{k_\mathrm{B}}          
 
\newcommand{\Rh}{R_\mathrm{h}}          
\newcommand{\ndens}{n}
\newcommand{\mdens}{\rho}

\newcommand{\uzh}{Department of Chemistry, Universit\"at Z\"urich, 8057 Z\"urich, Switzerland}
\newcommand{\ilm}{Univ Lyon, Univ Claude Bernard Lyon 1, CNRS, Institut Lumi\`ere Mati\`ere, F-69622, Villeurbanne, France}
\newcommand{\iuf}{Institut Universitaire de France (IUF), 1 rue Descartes, 75005 Paris, France}

\begin{document}


\title{Connection between water's dynamical and structural properties: insights from \textit{ab initio} simulations}

\author{Cecilia Herrero}
\affiliation{\ilm}
\author{Michela Pauletti}
\affiliation{\uzh}
\author{Gabriele Tocci}
\affiliation{\uzh}
\author{Marcella Iannuzzi}
\affiliation{\uzh}
\author{Laurent Joly}
\email{laurent.joly@univ-lyon1.fr}
\affiliation{\ilm}
\affiliation{\iuf}

\date{\today}

\begin{abstract}
Among all fluids, water has always been of special concern for scientists from a broad variety of research fields due to its rich behavior. In particular, some questions remain unanswered nowadays concerning the temperature dependence of bulk and interfacial transport properties of supercooled and liquid water, e.g. regarding the fundamentals of the violation of the Stokes-Einstein relation in the supercooled regime or the subtle relation between structure and dynamical properties. Here we investigated the temperature dependence of the bulk transport properties from \textit{ab initio} molecular dynamics based on density functional theory, down to the supercooled regime. We determined from a selection of functionals, that SCAN better describes the experimental viscosity and self-diffusion coefficient, although we found disagreements at the lowest temperatures. For a  limited set of temperatures, we also explored the role of nuclear quantum effects on water dynamics using \textit{ab initio} molecular dynamics that has been accelerated via a recently introduced machine learning approach. We then investigated the molecular mechanisms underlying the different functionals performance and assessed the validity of the Stokes-Einstein relation. We also explored the connection between structural properties and the transport coefficients, verifying the validity of the excess entropy scaling relations for all the functionals. These results pave the way to predict the transport coefficients from the radial distribution function, helping to develop better functionals. On this line, they indicate the importance of describing the long-range features of the radial distribution function.
\end{abstract}

\maketitle

Water is an ubiquitous liquid, essential for life on earth, and therefore constitutes one of the most important chemical substances.
Despite the apparent simplicity of its chemical formula, water is a complex liquid,
that after much effort, still evades our complete understanding at the
molecular level \cite{Gallo2016}. 
Due to its critical relevance with regard to energy harvesting and water purification, 
several efforts have been carried in order to obtain molecular insights about water behavior under different thermodynamic 
conditions. Water molecular interactions arise from a balance between van der Waals and hydrogen bonding forces \cite{Stillinger1980,Morawietz2016}, thus a complete description exclusively from classical force field (\ff) simulations may hinder some critical mechanisms. \textit{Ab initio} molecular dynamics (AIMD), 
where interatomic forces are computed from the electronic structure, 
may play a key role in understanding some important physical processes for bulk and confined water \cite{Lee2006,Cicero2008,Wohlfahrt2020,Ye2021} as, for instance, the controversial liquid-liquid critical point \cite{Gartner2020}.

A very efficient approach to determine the electronic structure is density functional theory (DFT), based on a formulation of the 
many-body problem in terms of
a functional of the electron density.
So far  however, many widely used approximations 
 for the exchange-correlation functional do not provide a sufficiently accurate description of many of water   properties \cite{Bankura2014,Gillan2016}. 
 The main challenge of semi-local and hybrid
 density functional approximations in predicting 
 the structure and energetics of water lies in
 their description of dispersion 
 and exchange-overlap interactions \cite{Gillan2016,DelBen2015}.
 Additionally, nuclear quantum effects (NQEs) play an important role in 
 determining water structure \cite{Ceriotti2016,Marsalek2017,Rossi2018,Ceriotti2009}: while
 on the one hand, NQEs tend to strengthen the hydrogen bond, 
on the other hand, competing effects 
 due to the spread of the protons
 in the normal direction tend to weaken it. Although NQEs 
 can be modelled via \textit{ab initio} path integral 
 molecular dynamics (PIMD), accounting for this subtle
 competition adds an additional
 level of complexity \cite{Ceriotti2016}.
 Despite recent advances in describing the water structure and 
 thermodynamic properties \cite{Cheng2019,reinhardt2021quantum},
 predicting transport properties from first principles represents a
 further challenge \cite{Kuhne2009,Chen2017,Lee2007}. 
Limited work has been dedicated to the study of the temperature
 evolution of the diffusivity and of the shear viscosity with \textit{ab initio} methods \cite{Gartner2020,Alfe1998,Morawietz2016}. 
A clearer picture of the molecular
mechanisms controlling the water viscosity and
diffusion is needed, 
especially in the supercooled regime \cite{Gallo1996,Debenedetti2003}, where water
viscoelasticity is poorly understood \cite{Schulz2020,AlmeidaRibeiro2020,oSullivan2019}
and the validity of the Stokes-Einstein (SE) relation at low temperatures
remains an open question \cite{Jung2004,Chen2006,Kumar2007,Xu2009,Shi2013a,Kawasaki2017}.

Water dynamics is also crucial in the field of nanofluidics \cite{Bocquet2010}, where in particular the performances can be boosted by liquid-solid slip, arising from a competition between bulk liquid viscosity and interfacial friction  \cite{Botan2011,Cross2018,Tocci2020,Herrero2020}.  
Further,
 reaching clearer insights into the
 molecular properties controlling water dynamics
 would enable to determine a relationship 
 between the structural correlations 
 and
 molecular transport. Establishing such a thermodynamic link between
 structure and dynamics would also be instrumental to improve the description of water via DFT.

Such connection has already been explored in the literature via \emph{e.g.} free-volume models \cite{Cohen1959,Turnbull1961}, relationships between $g(r)$ and glass transition temperature \cite{Ojovan2020}, and the proposition of different structural descriptors \cite{Russo2014,Shi2018origin,Tong2019}, among which the entropy excess scaling, already employed for AIMD simulations of liquid metals \cite{Gao2018} or water \cite{Zhang2011entropy}, stands out \cite{Dzugutov1996,Yokoyama1998,Ingebrigtsen2018}. The excess of entropy, which can be decomposed 
 in terms on the $N-$body radial distribution functions \cite{Baranyai1990}, has been proven to exhibit an exponential relation with the diffusion coefficient for multiple systems \cite{Fomin2014,Nandi2015}. In particular, for glass forming liquids such as supercooled binary mixtures and water, the approximation of the entropy excess by its two body contribution (related to an integral of a function of $g(r)$) has been shown to work well for a broad range of temperatures  \cite{Mittal2006,Mittal2007,Chopra2010,Ingebrigtsen2018,Bell2020}.
 One of the main limitations for AIMD is its great need of resources as compared to their classical counterparts. Nevertheless, if the link between dynamics and structure is established, we would be able to predict the transport coefficients from structural properties, which require shorter simulation times to converge \cite{Rotenberg2020}.
 Aside of this, entropy excess scaling has also been used as a tool to bring insights into the molecular mechanisms underlying 
 the SE relation \cite{Bell2020}.

In this report, we determine from a selection of density
functionals commonly used to characterise water \cite{DelBen2015,Gillan2016,Zheng2018},
which one better describes the temperature dependence of the
water viscosity and self-diffusion coefficient in its liquid and supercooled state, in comparison with \ff{} simulations using the TIP4P/2005 water model \cite{Abascal2005tip4p}. 
Additionally, we explore the connection between structural properties and the transport coefficients for all the functionals proposed via the two-body entropy parameter, presenting this physical descriptor as a path to develop better functionals and better compare with experimental results. 
We used AIMD simulations, describing the electronic structure within DFT, in the NVT ensemble to determine hydrodynamic bulk transport coefficients of three different density functionals: PBE \cite{Perdew1996} functional with Grimme's D3 corrections \cite{Grimme2004,Grimme2006} (namely \pbed), \optb{} \cite{Klimevs2009,Klimevs2011} and \scan{} \cite{Sun2015}. 
While describing the electronic structure with the SCAN functional, 
 we also included the role of NQEs by performing 
 PIMD simulations  
 for a limited set of temperatures,
 by employing a recently 
 introduced machine learning approach to speed-up
 the calculation of the electronic structure problem \cite{pauletti2021subsystem}.
Further simulation details can be found in \textit{Materials and Methods}.

\begin{figure}
\begin{subfigure}{0.49\linewidth}
    \centering
    \includegraphics[width=0.99\linewidth]{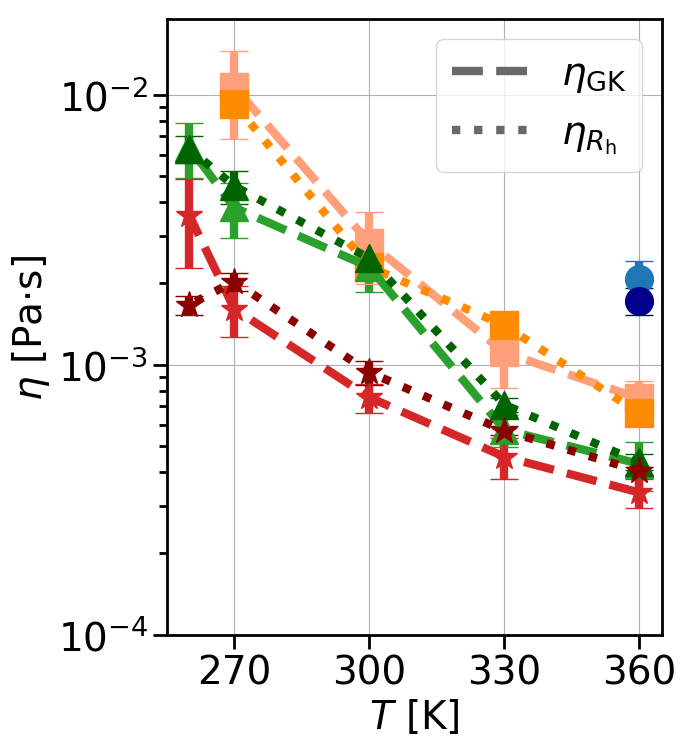}
    \caption{}
    \label{fig:etarh_funcs}
\end{subfigure}
\begin{subfigure}{0.49\linewidth}
    \centering
    \includegraphics[width=0.99\linewidth]{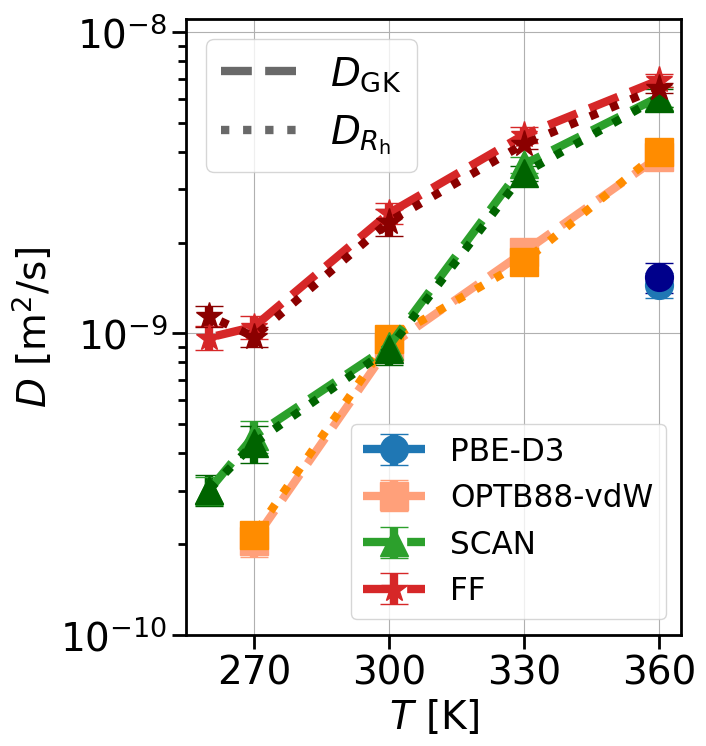}
    \caption{}
    \label{fig:Drh_funcs}
\end{subfigure}
\caption{Temperature evolution for different functionals of (a) shear viscosity 
and (b) diffusion coefficient. 
A good agreement is found between the hydrodynamic radius measures (dotted lines) and the Green-Kubo ones (dashed lines), implying all the functionals verify Stokes-Einstein relation with the same hydrodynamic radius $\Rh = 1\,$\AA{} (see text for detail).}
\label{fig:rh_funcs}
\end{figure}

\section*{Results and Discussion}
We display in Fig.~\ref{fig:etarh_funcs} (dashed lines) the temperature evolution of the shear viscosity, determined from the long-time plateau of the Green-Kubo integral, $\etaGK$ (Eq.~\eqref{eq:eta_gk} in Materials and methods), for the different functionals. No plateau was observed for \pbed{} at $T<360~$K and \optb{} at $T=260~$K. To benchmark the results, the same procedure was carried for \ff{} simulations with the TIP4P/2005 water model \cite{Abascal2005tip4p}, which provides an excellent description of experimental results for both viscosity and diffusion coefficient \cite{Tazi2012diffusion,Guillaud2017decoupling,Herrero2020}. In Fig.~\ref{fig:etarh_funcs} one can see that the viscosity obtained from the \scan{} functional is in better agreement with \ff{} at $330~$K and $360~$K, 
although between $260~$K and $300~$K it 
overestimates the viscosity by more than a factor of $1.7$.
With regard to \pbed{} and \optb{}, one observe from Fig.~\ref{fig:etarh_funcs} that both functionals overestimate 
$\etaGK$ value. Overall, all functionals fail at describing the temperature evolution of the shear viscosity.

The diffusion coefficient $\Dpbc$ was determined from the slope of the mean squared displacement in the diffusive regime (see the supporting information, SI). In practice, because of hydrodynamic interactions between the periodic image boxes, a finite size correction for the diffusion coefficient has to be introduced \cite{Yeh2004,Tazi2012diffusion,MonteroDeHijes2018}. For a cubic simulation box of size $\Lbox$ with periodic boundary conditions:
\begin{equation}
    \Dgk = \Dpbc + 2.837 \frac{\kB T}{6 \pi \etaGK \Lbox},
    \label{eq:Dcorrected}
\end{equation}
with $\kB$ the Boltzmann constant and $T$ the temperature. We denoted $\Dgk$ the diffusion coefficient obtained through Eq.~\eqref{eq:Dcorrected} because we used $\etaGK$ in it. $\Dpbc$ could not be determined within our simulation times for \pbed{} at $T<360~$K and \optb{} at $T=260~$K because the system did not enter in the diffusive regime. This result is consistent with the absence of a plateau for $\etaGK$. The corrected diffusion coefficient $\Dgk$ results are displayed in Fig.~\ref{fig:Drh_funcs} (dashed lines). In analogy to $\etaGK$, one observes in this figure that \scan{} is the functional that better describes water diffusion coefficient at high temperatures, although it fails at low $T$.

Generally, viscosity $\eta$ and diffusion $D$ are related through the SE relation:
\begin{equation}
    D = \frac{\kB T}{6 \pi \eta \Rh},
    \label{eq:SE}
\end{equation}
with $\Rh$ the effective hydrodynamic radius of the molecules \cite{Weiss2018}.
Even though the failure of this relation is well known at low temperatures \cite{Jung2004,Chen2006,Kumar2007,Xu2009,Shi2013a,Kawasaki2017}, it still remains valid for a broad range of temperature. We verified this statement by computing $\Rh$ for \ff{} simulations, obtaining constant $\Rh \sim 1~$\AA{} for the range of temperatures considered in the present study (see the SI).

Taking into account $D$ size correction Eq.~\eqref{eq:Dcorrected} and SE relation Eq.~\eqref{eq:SE}, one can relate the
viscosity to $\Dpbc$ and to the hydrodynamic radius:
\begin{equation}
    \etarh = \frac{\kB T}{6 \pi \Dpbc} \qty( \frac{1}{\Rh} - \frac{2.837}{\Lbox} ).
    \label{eq:etarh}
\end{equation}
In the same way, one can also determine a relation for $\Drh$ independent of $\eta$ from Eq.~\eqref{eq:Dcorrected} and Eq.~\eqref{eq:SE}:
\begin{equation}
    \Drh = \frac{\Dpbc}{1 - \frac{2.837 \Rh}{L_\mathrm{box}}}.
    \label{eq:Drh}
\end{equation}
Therefore, viscosity and diffusion can be determined exclusively from the slope of the mean squared displacement at long times by imposing the hydrodynamic radius $\Rh$. In order to test the applicability of this prediction, in Fig.~\ref{fig:rh_funcs} we display the results for $\etarh$ from Eq.~\eqref{eq:etarh} and $\Drh$ from Eq.~\eqref{eq:Drh} by imposing $\Rh=1\,$\AA{} (value in agreement with the \ff{} measures, see the SI). In Fig.~\ref{fig:rh_funcs} one can see a good match between the Green-Kubo and the hydrodynamic radius measures for both transport coefficients and for all the functionals considered, meaning that, although all the functionals fail in predicting viscosity and diffusion temperature dependence, all of them verify the SE relation 
with the same constant value of $\Rh$. 

\begin{figure}
    \centering
    \includegraphics[width=0.87\linewidth]{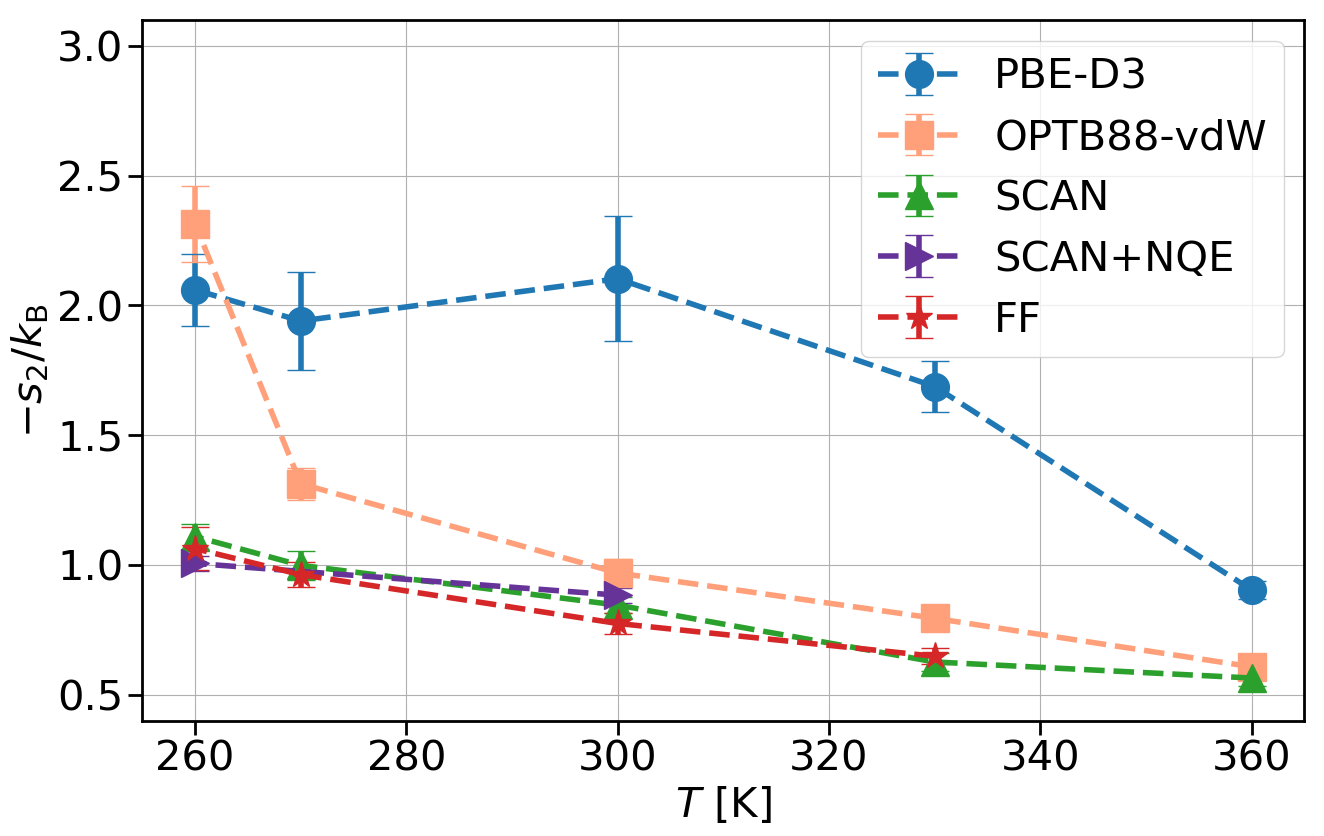}
    \caption{Dimensionless two-body entropy $s_2/\kB$ for different functionals and for \ff{} as a function of the temperature. }
    \label{fig:s2kb_temp}
\end{figure}

\begin{figure}
\begin{subfigure}{0.49\linewidth}
    \centering
    \includegraphics[width=0.99\linewidth]{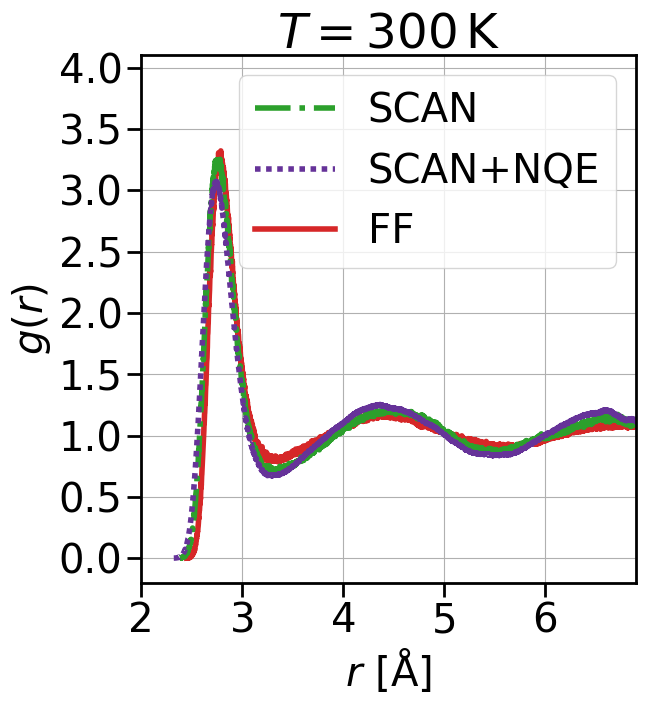}
    \caption{}
    \label{fig:rdf_nqe_T300}
\end{subfigure}
\begin{subfigure}{0.49\linewidth}
    \centering
    \includegraphics[width=0.99\linewidth]{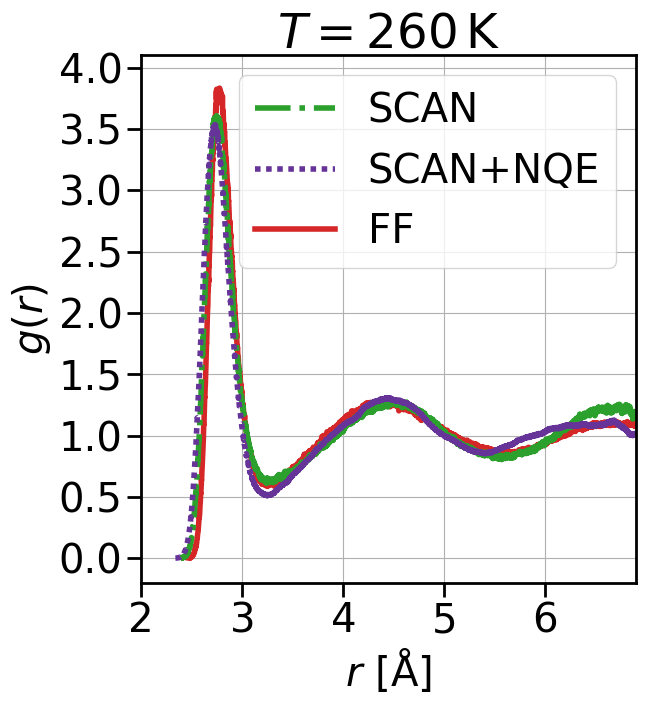}
    \caption{}
    \label{fig:rdf_nqe_T260}
\end{subfigure}\\
\begin{subfigure}{0.49\linewidth}
    \centering
    \includegraphics[width=0.99\linewidth]{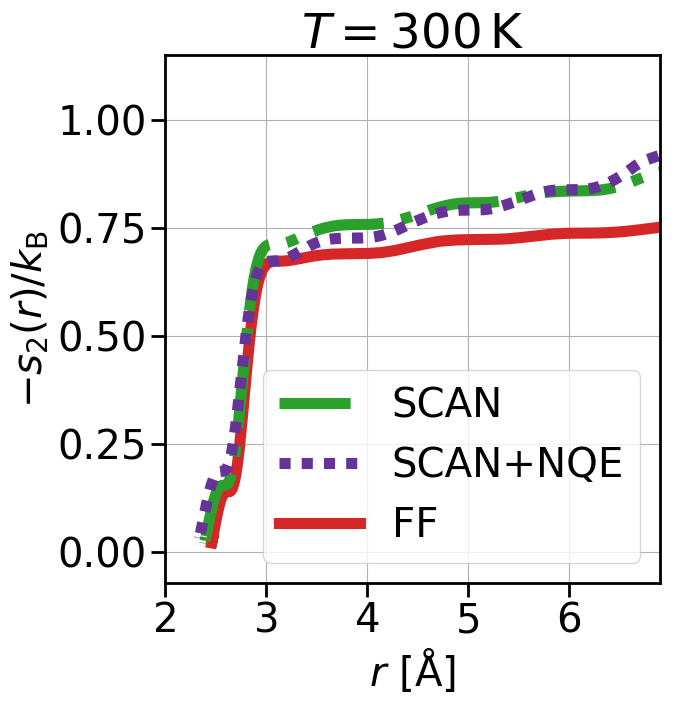}
    \caption{}
    \label{fig:s2kb_nqe_T300}
\end{subfigure}
\begin{subfigure}{0.49\linewidth}
    \centering
    \includegraphics[width=0.99\linewidth]{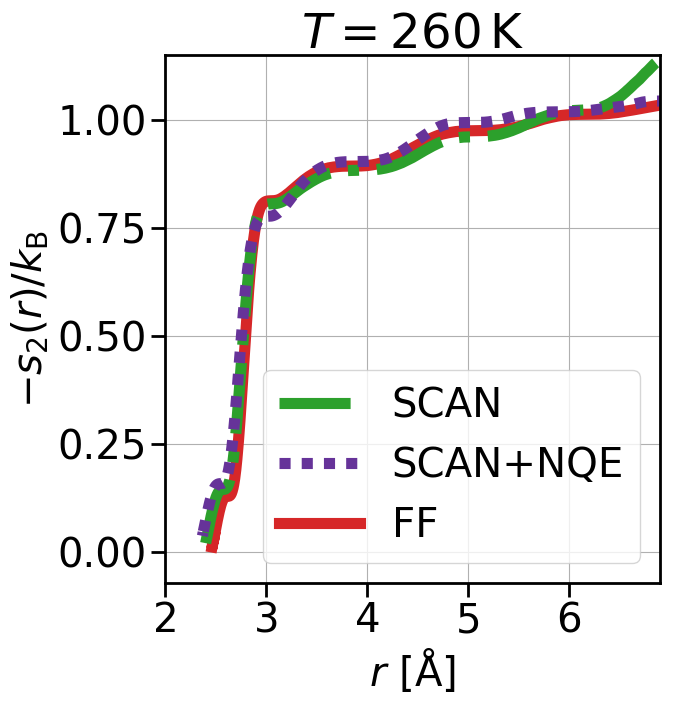}
    \caption{}
    \label{fig:s2kb_nqe_T260}
\end{subfigure}
\caption{Radial distribution functions (a) and (b) and $-s_2/\kB$
running integrals (c) and (c) of water at 300 K and 260 K obtained from
the \scan{} functional with classical
and quantum nuclei (\scan{}+NQEs) as 
well as from the \ff{} simulations.}
\label{fig:rdf_s2kb_nqe}
\end{figure}

Having determined the transport properties for the different functionals, we proceed to explore their connection with the structure of water, given by the radial distribution function, $g(r)$. Specifically, we computed the structural descriptor $s_2$ (two-body excess entropy, see the SI), given by the integral \cite{Baranyai1990}: 
\begin{equation}
    \frac{s_2}{\kB} = -2\pi \ndens \int_0^\infty r^2 \qty( g(r)\ln g(r) - g(r) +1 ) \, \dd r,
    \label{eq:s2_gr}
\end{equation}
with 
$\ndens$ the number density of the system.

Figure~\ref{fig:s2kb_temp} presents the temperature dependence of the dimensionless two-body entropy $s_2/\kB$ for the different functionals, compared with \ff{} results. 
We note that, although the 
$g(r)$ simulated with the TIP4P/2005 \ff{} exhibits
some discrepancies with respect to the experimental
$g(r)$ \cite{Abascal2005tip4p}, the change of $s_2$ with temperature 
is
in qualitative agreement with experiments for
a wide range of temperatures, down to the supercooled
regime \cite{Camisasca2019}. 

One can observe that \scan{} is the functional that better describes the $s_2$ temperature evolution, as compared to \ff. 
Interestingly, accounting for NQEs (see the SCAN+NQE in Fig.~\ref{fig:s2kb_temp})
does not produce significant changes in the
behavior of  $s_2/\kB$, 
even at the low temperature of $260$\,K. 
Whereas at the highest temperature of $360\,$K the \optb{} functional reproduces the water structure --
as represented by the $s_2$ order parameter, using this functional in the supercooled regime gives rise
to serious over structuring, thus leading
to an overestimation of more than two times
in the value of $s_2$ at $260\,$K.
Instead, \pbed{} fails at recovering the liquid two-body entropy at any temperature, and the integral in Eq.~\ref{eq:s2_gr} reaches a plateau for the lowest temperatures, \emph{i.e.} the $g(r)$ oscillations amplitude is not significantly affected by temperature, hinting at a possible glass transition \cite{Ojovan2020}.

To elucidate the connection between the water
structure and the transport coefficients through
the $s_2$ order parameter, we compare 
the radial distribution functions and the $s_2/\kB$ running
integrals for the \ff{} and the \scan{} functional
at 260\,K and at 300\,K, both with 
classical and with quantum
nuclei (see Fig.~\ref{fig:rdf_s2kb_nqe}),
while the comparison with the \pbed{} and
\optb{} functionals can be found in the SI.
At 300\,K the quantum nuclear $g(r)$ displays a
lower first peak compared to the classical 
one and to the $g(r)$ from the \ff{},
while the rise in the first peak also
occurs at a slightly shorter distance.
Thus, NQEs give rise to a less structured
first coordination shell, in qualitative
agreement with experimental results in H$_2$O and D$_2$O
which display these same characteristics, respectively \cite{soper2008quantum}.
Beyond the first peak, an increased structure
is observed in the $g(r)$ predicted with \scan{}+NQE,
compared to the case
with classical nuclei and with the \ff{} result,
in disagreement with experiments on light and heavy water \cite{soper2008quantum}.
Such discrepancy has been observed in previous
PIMD simulations obtained with semi-local
density functionals \cite{Ceriotti2016,gasparotto2016probing}, 
where it was pointed out that the origin of the increased
structure of the second peak arises from 
the destabilization of interstitial hydrogen
bonded configurations occurring in quantum nuclear
simulations, and more accurate descriptions might alter this balance
and lead to a less structured second and third 
solvation shells \cite{Cheng2019,del2015probing}.
Shifting the focus on the $s_2/\kB$
running integral (see
Fig.~\ref{fig:rdf_s2kb_nqe}(c)), it can be noticed that although
the largest contribution to the limiting value of $s_2/\kB$
arises from the first solvation shell, a non-negligible
part is also due to the oscillations beyond
the first solvation shell. Thus,
as a result of a less structured first
solvation shell and of more structured second and third shells,
the \scan{}+NQE functional predicts a 
limiting value of $s_2/\kB$ that is similar to
\scan{} with classical nuclei,
whereas the $s_2/\kB$ value predicted with \ff{}
is visibly lower.
Interestingly, the differences between the quantum
and classical nuclear $g(r)$, observed at $300$\,K,
are attenuated at $260\,$K, also when
comparing with the \ff{} results.
This leads to a value of $s_2/\kB$ that
is remarkably similar
for all the three methods at $260\,$K, 
as seen in Fig.~\ref{fig:rdf_s2kb_nqe}(d).

Finally, in order to establish a relation between structure and transport coefficients, we proceed to test the entropy excess scaling laws. With that regard, it has been verified that the entropy excess $s_\mathrm{ex}$ can be approximated by the two-body contribution $s_\mathrm{ex} \simeq s_2$ for water and supercooled binary mixtures \cite{Mittal2006,Mittal2007,Chopra2010,Bell2020}. 
As described in Ref.~\citenum{Agarwal2011}, $s_2$ is constructed from the oxygen-oxygen,
oxygen-hydrogen, and hydrogen-hydrogen pair distributions. Still, the scaling laws hold well by just computing $s_2$ from the oxygen-oxygen $g(r)$, amounting to only considering the translational 2-body entropy, with a difference of a factor 3 between both estimates, so $s_\mathrm{ex} \simeq 3 s_2$ from our results.
It can be shown \cite{Dyre2018} (see the SI) that the dimensionless diffusion coefficient $D/D_0$ is expected to scale as:
\begin{equation}
    \frac{D}{D_0} = A \exp(-B\, s_2/\kB),
    \label{eq:DD0_s2}
\end{equation}
with $D_0 = l_0 \sqrt{\kB T/m}$ (where $l_0 = n^{-1/3}$ is the average interparticle distance and $m$ is the fluid mass), and $A$ and $B$ dimensionless constants at a given fluid density. Considering this Eq.~\eqref{eq:DD0_s2}, together with the SE equation, and assuming $\Rh \sim l_0$, one can expect a scaling for the dimensionless viscosity $\eta/\eta_0$ as:
\begin{equation}
    \frac{\eta}{\eta_0} = A' \exp(-B' \, s_2/\kB),
    \label{eq:etaeta0_s2}
\end{equation}
with $\eta_0=\sqrt{m \kB T}/l_0^2$. From the SE relation one expects $B'=-B$.

\begin{figure}
\begin{subfigure}{0.49\linewidth}
    \centering
    \includegraphics[width=0.99\linewidth]{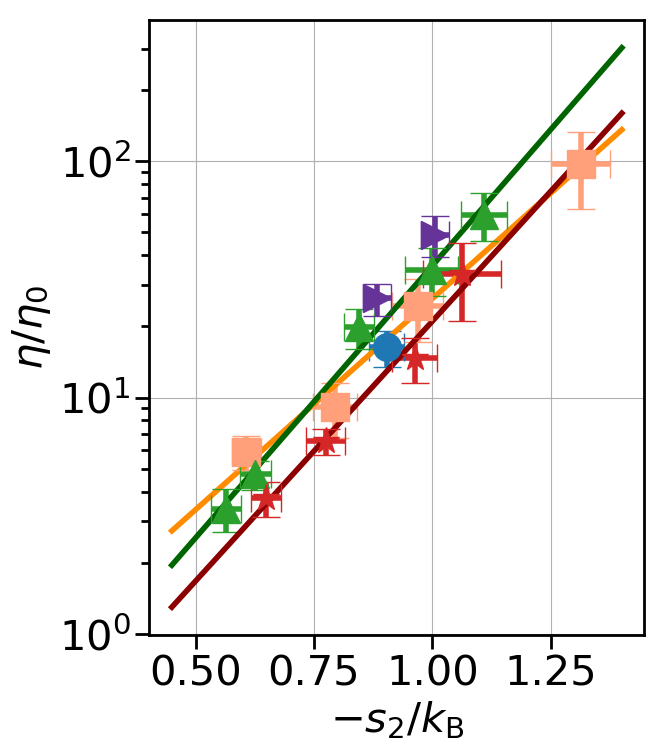}
    \caption{}
    \label{fig:etaeta0_s2kb}
\end{subfigure}
\begin{subfigure}{0.49\linewidth}
    \centering
    \includegraphics[width=0.99\linewidth]{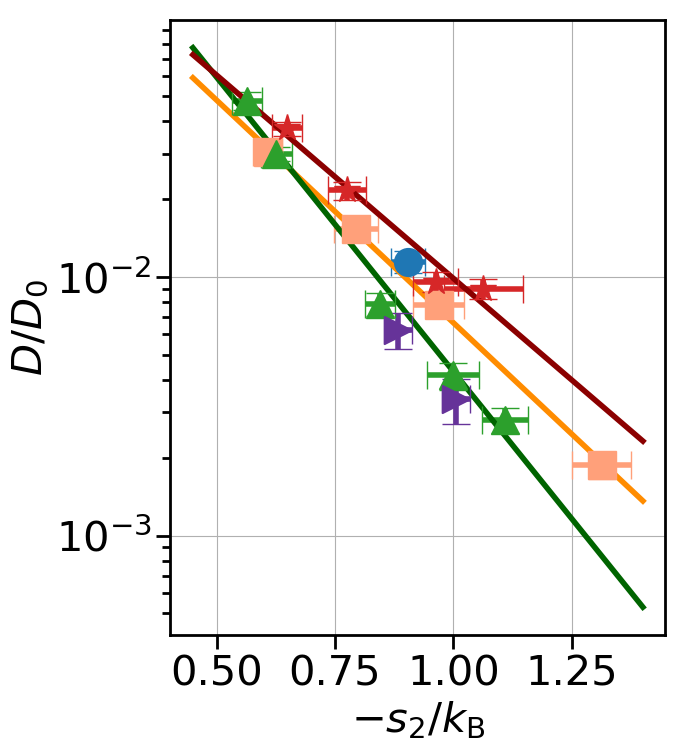}
    \caption{}
    \label{fig:DD0_s2kb}
\end{subfigure}
\caption{Reduced (a) viscosity $\eta/\eta_0$ and (b) diffusion coefficient $D/D_0$, defined in Eq.~\eqref{eq:DD0_s2} and Eq.~\eqref{eq:etaeta0_s2}, as a function of the dimensionless two-body entropy $s_2/\kB$ for different functionals and \ff{} simulations. In continuum line are represented the respective exponential fits for each functional. The fit results are detailed in Table~\ref{tab:fitetaD_s2}. The color and marker style representing the different functionals is the same as in Fig.~\ref{fig:s2kb_temp}.}
\label{fig:xx0_s2kb}
\end{figure}

\begin{figure}
\begin{subfigure}{0.49\linewidth}
    \centering
    \includegraphics[width=0.99\linewidth]{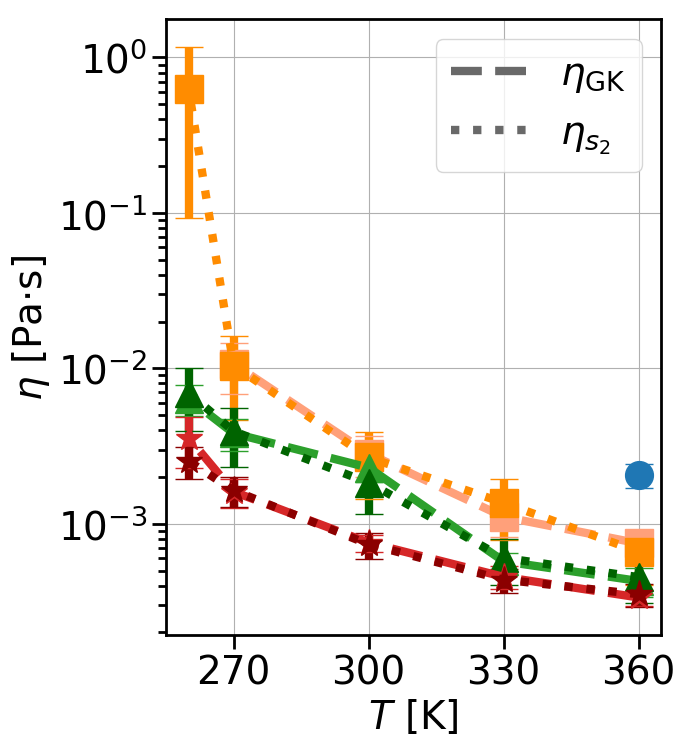}
    \caption{}
    \label{fig:etas2_funcs}
\end{subfigure}
\begin{subfigure}{0.49\linewidth}
    \centering
    \includegraphics[width=0.99\linewidth]{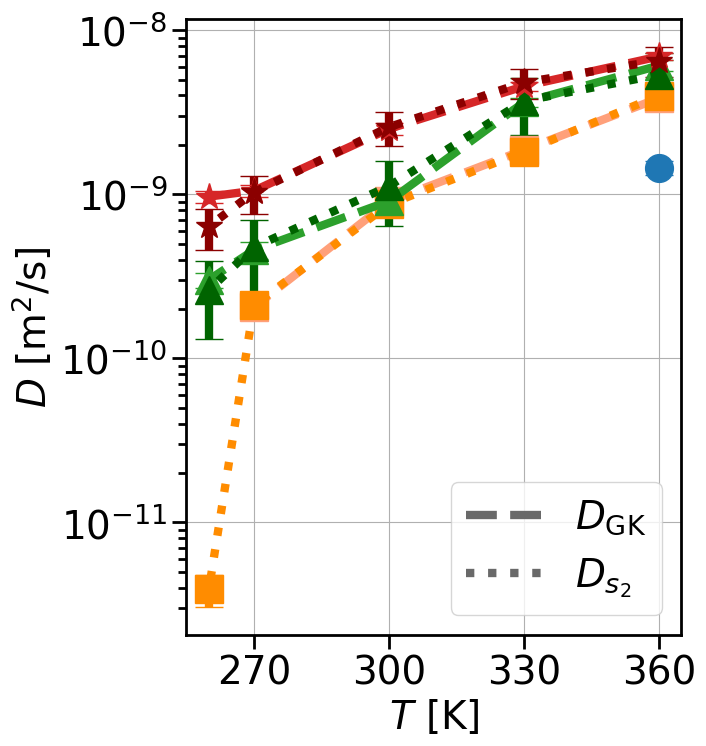}
    \caption{}
    \label{fig:Ds2_funcs}
\end{subfigure}
\caption{Temperature evolution for different functionals of (a) shear viscosity from Eq.~\eqref{eq:etaeta0_s2} and (b) diffusion coefficient from Eq.~\eqref{eq:DD0_s2} with fit parameters from Table~\ref{tab:fitetaD_s2}. A good agreement is found between the $s_2$ based prediction (dotted lines) and the explicit calculation (dashed lines), verifying the link between the structure and the transport coefficients. The color and marker style representing the different functionals is the same as in Fig.~\ref{fig:rh_funcs}.}
\label{fig:s2_funcs}
\end{figure}

We tested Eq.~\eqref{eq:DD0_s2} and Eq.~\eqref{eq:etaeta0_s2} for the different functionals. Figure~\ref{fig:xx0_s2kb} shows the results for the dimensionless transport coefficients as a function of the two-body entropy excess for the different functionals. One can see that, although the functionals predict different transport coefficients (Fig.~\ref{fig:rh_funcs}) and $s_2$ results (Fig.~\ref{fig:s2_funcs}), all of them verify an exponential scaling of $\etaGK/\eta_0$ and $\Dgk/D_0$ with $s_2$. Therefore, we fitted the relations in Eq.~\eqref{eq:DD0_s2} and Eq.~\eqref{eq:etaeta0_s2} for \optb, \scan, and \ff{} (continuous lines in Fig.~\ref{fig:xx0_s2kb}). No fit was performed for \pbed{} due to the single value measure we could report for this functional. The fit results are indicated in Table~\ref{tab:fitetaD_s2}. One can observe that, although out of the error bars, the fit parameters for \scan{} and \ff{} are the closest ones and that, for all functionals, $B'=-B$, implying a verification of the SE relation, Eq.~\eqref{eq:SE}.

\begin{table}
\centering
\caption{Fit parameters of the two-body excess entropy scaling relation for the dimensionless viscosity and diffusion coefficient, for different functionals and \ff{} simulations. The fit corresponds to the function $y = A\exp(-B s_2/\kB)$ with $y$ the dimensionless viscosity $\etaGK/\eta_0$, following Eq.~\eqref{eq:etaeta0_s2}, and diffusion coefficient $\Dgk/D_0$, following Eq.~\eqref{eq:DD0_s2}.} 
\label{tab:fitetaD_s2}
\begin{tabular}{lcccc}
\cmidrule{2-5}
& \multicolumn{2}{c}{ $\etaGK/\eta_0$ } & \multicolumn{2}{c}{$\Dgk/D_0$} \\
\cmidrule{2-5}
  & $A'$  & $B'$ &  $A $ & $B$ \\
  & $(\times 10^{-1})$ & & $(\times 10^{-1})$ & \\
\midrule
 \optb  &  $4.29(1.39)$ & $4.11(0.34)$ & $ 3.52(0.29)$ & $-3.97(0.09)$\\
 \scan  &  $1.79(0.48)$ & $5.31(0.31)$ & $ 8.17(2.49)$ & $-5.24(0.36)$\\
 \ff    &  $1.92(0.26)$ & $4.52(0.18)$ & $ 7.73(1.25)$ & $-4.58(0.21)$\\ 
\bottomrule
\end{tabular}

\end{table}

One can exploit the exponential relationship between the
transport coefficients and $s_2$ to predict transport properties 
from structural ones: once the fitting parameters in
Eqs.~(\ref{eq:DD0_s2}-\ref{eq:etaeta0_s2})
have been extracted by calculating the
dependence of $\eta$ and
$D$ on $s_2$ for a limited set of temperatures,
the value of the transport
coefficients can be obtained just from
the calculation of the $s_2$ order parameter
via the radial distribution function also for a wider
temperature range.
Indeed, generally structural
properties such as the $g(r)$ require shorter simulations to converge, especially when using force based methods, as the one proposed in Ref.~\citenum{Rotenberg2020}, to reduce the variance when compared to the conventional strategies based on particles positions binning. Figure~\ref{fig:s2_funcs} presents the Green-Kubo results and their comparison with the prediction resulting from the fit via $s_2$. One can see good agreement between the explicit calculation of transport coefficients and their and predictions via $s_2$ for all the data. Also, 
although the transport coefficients could not be calculated explicitly
for the \optb functional at the lowest temperature of $260$\,K,
they could be determined from the exponential relationship with $s_2$,
yielding an exceedingly high viscosity and low diffusion,
and thus verifying the failure of
this functional in reproducing the temperature
dependence of both transport coefficients.

\section*{Conclusions}

In summary, among the selected DFT functionals,
\scan{} best captures the temperature 
evolution of water transport properties -- as described by the the accurate TIP4P/2005 \ff{} -- despite
the disagreement observed at temperatures of 300\,K and below. 
We detected large discrepancies between functionals, with a major failure of \pbed, which is far too viscous. 
Despite these discrepancies, 
the SE relation was observed to hold in the considered temperature range for all the functionals and,
moreover, all of them predicted the same hydrodynamic radius $\Rh \sim 1~$\AA. This property, together with the finite size correction for the diffusion coefficient, allowed us to propose a measure of viscosity $\etarh$ and diffusion coefficient $\Drh$, based only on the slope of the mean squared displacement in the diffusive regime $\Dpbc$, for known box size and fixed $\Rh$.

Motivated by a possible connection between dynamics and structure, we computed the radial distribution functions for the different functionals. Analogously to the transport coefficient results, we observed that \scan{} radial distribution function is the one that better compares to \ff{}, with little differences at the lowest temperatures in the second and third solvation shells, whereas \pbed{} was far more structured that \scan{} and \optb{} at high temperatures, in agreement with the high viscosity value measured for this functional. An explicit relationship between dynamic and structure can be established through the two-body entropy excess, which is an integral of a function of the radial distribution function. We verified the exponential relation on $s_2$ for both reduced bulk transport coefficients and, although the connection between $s_2$
and transport properties is not univocal, as different 
fitting parameters were used for each functional,
the fitting parameters all have the same order of magnitude. 
We suggest that the non-universality of the exponential relation is due to the use of the translational two-body excess entropy; a more universal relation could be observed by using the full two-body excess entropy, although the related structural features would be less easy to interpret. 

Finally, based on the established exponential relation between the bulk transport coefficients, we computed both viscosity and diffusion coefficient from the $s_2$ results and the fitting parameters. This allowed us to estimate transport coefficients for functionals strongly structured (for instance \optb{} at $260\,$K), which present such a high viscosity value that longer simulations are needed in order to observe a well-defined plateau in the Green-Kubo integral. Therefore we propose here
that, once the exponential dependence has been determined for a few temperatures, the viscosity and diffusion coefficient can be determined only from structural properties, which typically require shorter simulation times to converge \cite{Rotenberg2020}. This can be a useful technique to apply in order to calculate transport coefficients for very viscous systems, where the associated time-scales are far from the ones computationally reachable with AIMD simulations.
The connection between transport coefficients and the radial distribution function via the two-body entropy excess also establishes some guidelines to choose a functional for simulations of nanofluidic systems, where a functional which better reproduces water's structure will more likely reproduce its dynamical properties. The $s_2$ order
parameter can be also employed as an instrumental tool
to 
gauge the potential of DFT or of high accuracy methods in describing dynamics without computing transport properties explicitly, 
where the comparison between different $s_2$ values becomes more straightforward than the comparison between two $g(r)$ profiles, or just the value of the $g(r)$ minimum or maximum, which does not ensure a full structure correlation. Indeed, from the $s_2(r)$ running integrals, we showed the importance of reproducing not only the first solvation shell of the $g(r)$ but also the long range structure, which is a non-negligible contribution to $s_2$ value. This feature, together with the scaling behavior of the bulk transport coefficients
as a function of entropy, suggests that it is important
that DFT reproduces not only the first peak in the $g(r)$,
but also its long range behavior, which is
critical to obtain an accurate description
of dynamical properties such as viscosity and diffusion coefficient.

It is worth discussing the possible
origins of the discrepancies of the temperature 
evolution of of the viscosity and of the 
diffusion coefficient, especially striking for 
the \scan{} functional, which shows good 
agreement with FF and 
experiments at high temperatures
but it overestimates the 
viscosity and underestimates the diffusivity at 
low temperatures. 
Although one might expect that the
inclusion of zero-point energy and quantum
tunneling, which become increasingly relevant at lower temperatures, would
play an important role, we have shown that taking NQEs into account did not improve upon the results obtained with classical nuclei.
As such, the most likely source of discrepancy lies in the approximate 
description of the electronic structure with
the chosen functionals.
Capturing the delicate 
balance between van der Waals
dispersion and exchange interactions 
constitutes the main challenge for the
description of water \cite{Gillan2016}
and it is also critical in order to predict
water transport properties below
room temperature. It remains to be seen
whether the use of high-accuracy 
methods such as the random-phase 
approximation (RPA), M{\o}ller–Plesset
perturbation theory \cite{DelBen2015} and
quantum Monte Carlo \cite{zen2015ab}
would improve upon the 
current description of water
transport properties also in super-cooled
conditions. Recent results on the
diffusion coefficient for a wide range of temperatures obtained with RPA, also including
the role of NQEs are promising in this regard \cite{yao2021nuclear}. As a further interesting perspective of this work, the established connection between the
structure and dynamics might reveal
what tips the balance between 
the strengthening and the weakening
quantum delocalization effects of the H-bond
network in water.
Since diffusion is found experimentally to vary significantly between D$_2$O and H$_2$O \cite{Ceriotti2016},
and $s_2$ is directly connected
to diffusion, one can gauge the impact of
the competing H-bond
strengthening and weakening
NQEs on the structure and the
dynamics directly from the experimental
measurements of the diffusivity
and of the radial distribution function
of light and heavy water at different
temperatures.

\section*{Materials and methods}

\subsection*{Simulation Details}
We performed AIMD simulations of 32 water molecules in bulk using DFT with the CP2K code (development version) \cite{kuhne2020cp2k}, which employs the Gaussian and Plane waves (GPW) method
to describe the wave-function and the electron
density and to solve the Kohn-Sham equations \cite{VandeVondele2005}. Three different density functionals were considered: PBE \cite{Perdew1996} functional with Grimme's D3 corrections \cite{Grimme2004,Grimme2006}, \optb{} \cite{Klimevs2009,Klimevs2011} and \scan{} \cite{Sun2015}. The electronic structure problem was solved within the Born-Oppenheimer approximation for 5 different temperatures ($T=\{260,270,300,330,360\}\,$K (the two lowest ones corresponding to the expected supercooled regime) controlled via the Nos\'e-Hoover thermostat. We worked at constant volume with a box size such that $\mdens = 1\,$g/cm$^3$ ($L_\mathrm{box}=9.85\,$\AA{} for 32 water molecules). The running time was $\simeq 120\,$ps for all functionals and temperatures except \optb{} and $T=\{260,270\}\,$K, with running time $\simeq 240\,$ps. The timestep considered was $0.5\,$ps. The initial configuration for all the functionals corresponded to the steady state at the given temperature obtained from force field (FF) MD after a running time of $200\,$ps. The energy cutoff for plane waves was $600\,$Ry for \pbed{} and \optb{}, and $800\,$Ry for \scan{}, and the localized Gaussian basis set was short range molecularly optimized double-$\zeta$ valence polarized (DZVP-MOLOPT-SR) \cite{VandeVondele2007}. 

NQEs were modelled using PIMD simulations with a
thermostatted ring polymer contraction scheme
using 24 replicas \cite{craig2004quantum,habershon2013ring}. 
PIMD simulations were performed
using the i-PI code \cite{kapil2019pi} together with CP2K,
where the former is used
to propagate the quantum nuclear dynamics,
whereas the latter is used
for the optimization of the wave-function
and to calculate the forces on each atom.
Sampling in the canonical ensemble 
has been carried out at 260\,K and at 300\,K 
for 35\,ps.
Further, in the case of the PIMD simulations,
the electronic structure problem is solved
using subsystem density functional theory
within the Kim-Gordon (KG) scheme, where the
shortcomings of the electronic kinetic energy
term of KG-DFT are addressed by correcting
this term via a $\Delta$-machine learning approach \cite{ramakrishnan2015big}.
Specifically, we use a neural-network
potential based on the Behler-Parrinello 
scheme \cite{behler2007generalized} 
to learn the difference in the total energy
and atomic forces between KS-DFT and KG-DFT
(see Ref. \citenum{pauletti2021subsystem} 
and the SI for further
details). The resulting $\Delta$-learning
method provides the same accuracy
of KS-DFT at the lower 
computational cost of KG-DFT.

We also performed force field (classical MD) simulations via the LAMMPS package \cite{Thompson2021lammps}. Analogously to AIMD, we worked in the NVT ensemble with the temperature controlled via a Nos\'e-Hoover thermostat and with a volume such that $\mdens=1\,$g/cm$^3$. Three different box sizes were considered: $32$ water molecules ($L_\mathrm{box}=9.85\,$\AA), $64$ water molecules ($L_\mathrm{box}=12.42\,$\AA) and $128$ water molecules ($L_\mathrm{box}=15.64\,$\AA). The water model considered in all the cases was TIP4P/2005 \cite{Abascal2005tip4p}.

\subsection*{Shear Viscosity and Diffusion Coefficient}
For all the simulations, we determined the shear viscosity from the Green-Kubo relation:
\begin{equation}
    \etaGK = \frac{V}{\kB T}\int_0^\infty \expval{p_{ij}(t)p_{ij}(0)} \dd t,
    \label{eq:eta_gk}
\end{equation}
with $V$ the volume, $\kB$ the Boltzmann constant, $T$ the temperature and $p_{ij}=\{p_{xy},p_{xz},p_{yz}\}$ the non-diagonal components of the stress tensor. 

The error bars were computed within $60\%$ of confidence level. For viscosity, the total MD stress was divided in three time slots of equal length, each of them containing three independent measures of $\eta$. $\etaGK$ was measured at the time where the $\eta(t)$ running integral reached a time plateau (see the SI). Therefore 9 independent viscosity values were computed for each functional at a given temperature. For the diffusion coefficient, the first $20\,$ps were removed from the trajectory so the system equilibration from the initial configuration does not affect the mean squared displacement results. From them, 3 independent measures of $\Dpbc$ were obtained from the three independent Cartesian components.

\begin{acknowledgments}
The authors thank G. Galliero, S. Gelin and J.-M. Simon for fruitful discussions. 
We are also grateful for HPC resources
from GENCI/TGCC (grants A0050810637 and A0070810637),  
from the PSMN mesocenter in Lyon 
and from the Swiss National Supercomputing Centre (CSCS)
under project ID uzh1.
LJ is supported by the Institut Universitaire de France. GT   is   supported   by   the   Swiss National Science Foundation through the  project PZ00P2\_179964.
\end{acknowledgments}


%

\end{document}